\documentclass[letterpaper]{sig-alternate-2013}

\permission{\copyright 2017 International World Wide Web Conference Committee (IW3C2), 
published under Creative Commons CC BY 4.0 License.}
\conferenceinfo{WWW 2017,}{April 3--7, 2017, Perth, Australia.}
\copyrightetc{ACM \the\acmcopyr}
\crdata{ 978-1-4503-4913-0/17/04. \\
http://dx.doi.org/10.1145/3038912.3052684 \\
\includegraphics{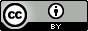}
}
 
\usepackage{color, colortbl}
\definecolor{lightgreen}{RGB}{152,251,152}
\definecolor{lightred}{RGB}{205,92,92}
\definecolor{lightyellow}{RGB}{238,232,170}

\usepackage{catchfile}
\usepackage{booktabs}
\usepackage{graphicx}
\usepackage{subcaption}
\usepackage{array,multirow}
\usepackage{wrapfig}
\usepackage{xcolor}
\usepackage{hyperref}
\usepackage{times} %

\usepackage{xspace}

\newcommand{\communityname}[1]{{\small\sf #1}\xspace}
\renewcommand\paragraph[1]{\noindent\textbf{\emph{#1}}}

\newcommand{\fuser}{user\xspace}
\newcommand{\fUser}{User\xspace}

\newcolumntype{L}[1]{>{\raggedright\let\newline\\\arraybackslash\hspace{0pt}}m{#1}}
\newcolumntype{C}[1]{>{\centering\let\newline\\\arraybackslash\hspace{0pt}}m{#1}}

\begin{document}

\title{Cats and Captions vs. Creators and the Clock:\\
  Comparing Multimodal Content to Context in Predicting Relative Popularity}

\numberofauthors{3}

\author{
\alignauthor
Jack Hessel \\
\affaddr{Dept. of Computer Science}\\
       \affaddr{Cornell University}\\
       \email{jhessel@cs.cornell.edu} \\
\alignauthor
Lillian Lee\\
       \affaddr{Dept. of Computer Science}\\
       \affaddr{Cornell University}\\
       \email{llee@cs.cornell.edu} \\
\alignauthor
David Mimno \\
       \affaddr{Dept. of Information Science}\\
       \affaddr{Cornell University}\\
       \email{mimno@cornell.edu} \\
}

\newif{\ifhidecomments}
\hidecommentsfalse
\hidecommentstrue

\ifhidecomments
\newcommand{\llee}[1]{}
\newcommand{\david}[1]{}
\newcommand{\jack}[1]{}
\else
\newcommand{\david}[1]{\textcolor{blue}{#1}}
\newcommand{\llee}[1]{\textcolor{magenta}{#1}}
\newcommand{\jack}[1]{\textcolor{red}{#1}}
\fi

\maketitle
\begin{abstract}
The content of today's social media is becoming more and more
rich,
increasingly mixing text, images, videos, and audio.
It is an intriguing research question to model the interplay between
these different modes in attracting user attention and engagement.
But in order to pursue this study of multimodal content, we must also
account for context: timing effects, community preferences,
and social factors (e.g., which authors are already popular)
also affect the amount of
feedback and reaction that social-media posts receive.  In this work,
we separate out the influence of these non-content factors in several
ways.  First, we
focus on ranking pairs of submissions posted to the same community in
quick succession, e.g., within 30 seconds;
this framing encourages models to focus on time-agnostic and
community-specific content features.
Within that setting, we determine the relative performance of author
vs. content features.  We find that victory usually
belongs to ``cats and captions,'' as visual and textual features
together tend to outperform identity-based
features.  Moreover, our experiments show that when considered in
isolation, simple unigram text features and deep neural network visual
features yield the highest accuracy individually, and that the
combination of the two modalities generally leads to the best
accuracies
overall.
\end{abstract}


\interfootnotelinepenalty=10000
\section{Introduction}

\begin{figure}
\centering
\includegraphics[width=.45\textwidth]{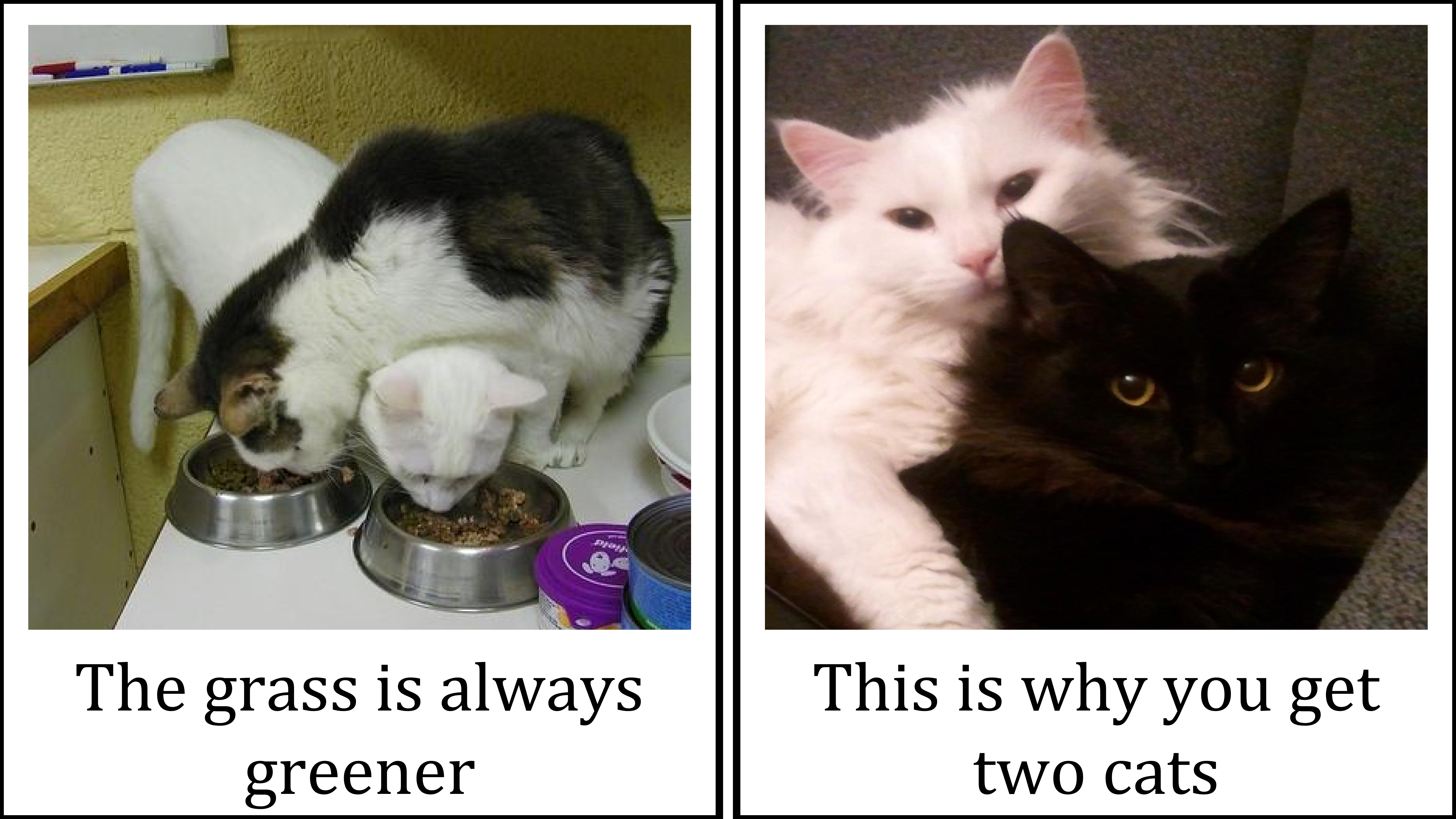}
\caption{Despite being submitted only 13 seconds apart to the
  subreddit \communityname{aww}, one of these submissions received
  over 1600 upvotes whereas the other received fewer than 20; the
  answer is in Section 3. Images courtesy \texttt{imgur.com}, posted
  by Reddit users mercurycloud and imsozzy.}
\label{fig:cats}
\end{figure}

Today's user-generated content is becoming more multimodal as users
increasingly mix text, images, videos, and audio.  Does one mode tend
to be preferred over another --- for example, on the Internet, is it
indeed true that ``a picture is worth a thousand words''?  Or do the
visual and the linguistic interact, sometimes reinforcing and
sometimes counteracting each other's individual influence?
Anecdotally, at least, it seems that there is interesting interplay
between these different modes.  For example, Figure \ref{fig:cats}
compares two posts made to the same forum on the same site, both
containing captioned images of two cats. One could argue that the
leftmost one has a more clever caption\footnote{One user comments in
response: ``A good title! Refreshing. Better than `this lil guy.'''}
but the second has a more attractive image. Which would more users
prefer?

However, determining what multimodal content is most attractive is
complicated by the fact that popularity can be strongly dependent on
many non-content
factors \cite{suh2010want,Bakshy:ProceedingsOfWsdm:2011,hong2011predicting,ma2012will,borghol2012untold,Romero+Tan+Ugander:13,lakkaraju2013s}.
Posts by users that already have a large audience tend to enjoy an
advantage over posts by relatively unknown people; posts that appear
when users are most active are also more popular; and sometimes simply
the fact that a post receives a few early clicks ensures that it gains
even more popularity.

Yet to dismiss the importance of the {\em content} of a post would be
wrong.  From a user's perspective, if content matters less than
identity and timing, why would they bother taking better pictures or
writing wittier captions?  Community moderators, who would ideally
like to promote high quality content even if it was submitted at a
less-than-optimal time or by a non-celebrity user, would also
appreciate a model of content alone. Researchers trying to understand
community preferences/biases want to model users' likes and dislikes,
not the idiosyncrasies of ranking algorithms and random early upvoting
patterns.

In this work, we seek to measure content preferences
independent
of confounding factors.
We collect and analyze data from six sub-communities on {\tt reddit.com}
of varying size and focus.  Each focuses on {\em multimodal} posts
that include images and captions.
Inspired by our prior work on wording effects \cite{danescu2012you,tan2014effect}, we
select pairs of captioned images posted at \emph{similar times} (e.g.,
30 seconds) to the \emph{same community} and then construct models to predict which of the two
eventually becomes more popular. Comparing submissions in this
time-controlled setting
allows us to approach an ``equal footing'' assumption when modeling
content, and to quantify the validity of that assumption.

We choose to explicitly control for time of posting because we find
that it is the most important contextual factor and because it is
relatively easy to find comparable pairs.  But there are other factors
in play, some Reddit-specific and that are impossible for us to
recover, e.g., the precise ordering of content displayed to users.
However, we perform human annotation experiments that verify that
these unrecoverable factors do not overwhelm the influence of content
(see \S \ref{sec:eda} for more details).  For other factors, such as a
user's social status or experience, we take the approach of
quantifying the predictive performance of such effects relative to
content features, since explicitly controlling for both timing and
user would leave us with too little data to work with.

When comparing ``cats and captions'' --- that is, post content --- to
creator characteristics,
we find that ``cats and captions'' are generally more important for
the communities we examine. Also, while image features always
outperform text features when both are considered independently
(albeit only if deep learning is employed), in five of six Reddit
communities, significant performance gains are observed when combining
modalities.

The main contributions of this work are:
\begin{enumerate}
\item An exploration of time-sensitive content popularity across
  various communities on \texttt{reddit.com}, and an accompanying
  argument for framing these investigations in a time-controlled, ranking
  setting.
\item Several publicly available\footnote{\href{www.cs.cornell.edu/~jhessel/cats/cats.html}{\texttt{www.cs.cornell.edu/\textasciitilde{}jhessel/cats/cats.html}}} datasets and ranking tasks involving
  the prediction of community response to multimodal content, plus
  estimates of human performance on these tasks.
\item A comparison of off-the-shelf image and language features
  against social and timing baselines, and a demonstration that
  multimodal features are worth incorporating. The models we consider
  can also be applied to submissions in isolation, enabling on-line
  scoring of novel content.
\end{enumerate}

\section{Datasets}

Our starting point for Reddit data is Tan and Lee's
\shortcite{tan+lee:15} dataset of all 106M submissions to Reddit from
2007 to 2014 and Hessel et al.'s \shortcite{hessel2016science}
extension of this dataset to include full Reddit comment
trees. Reddit, which is the $25^{th}$ most popular site on the
Internet according to \texttt{Alexa.com} as of Fall 2016, consists of
interest-centric subcommunities called subreddits. These datasets are
based on the work of Jason Baumgartner of \texttt{pushshift.io} who
scraped Reddit using their public API.

On Reddit, users are allowed to up/downvote content submitted by other
users. While the exact counts of each of these votes are not made
available,\footnote{The exact totals are obscured to prevent spam.}
Reddit computes and displays a proprietary ``engagement'' metric based
on the number of upvotes minus the number of downvotes. This quantity,
called the \emph{score} of a post, has been readily used in previous
work, and is the measure of engagement we will be examining.

Content on Reddit is shared with topical subreddits (e.g.,
\communityname{politics}, \communityname{Art}); this allows us to
control for the types of content by only comparing submissions within
a given subreddit. In contrast to a majority of previous work that
uses general-purpose image datasets from Flickr for popularity
prediction, we examine a wide variety of \emph{granularities} of
content, ranging from highly general to very fine. Khosla et
al. \cite{khosla2014makes}, for example, find that objects like
revolvers and women's bathing suits are predictive of popularity,
whereas spatulas, plungers, and laptops have a negative impact.
In other words, while previous work has addressed which types of
objects tend to become popular, here we examine what objects of a
\emph{given} type become popular.

\begin{figure}
\centering
\begin{minipage}{.22\textwidth}
  \centering
  \includegraphics[width=\textwidth]{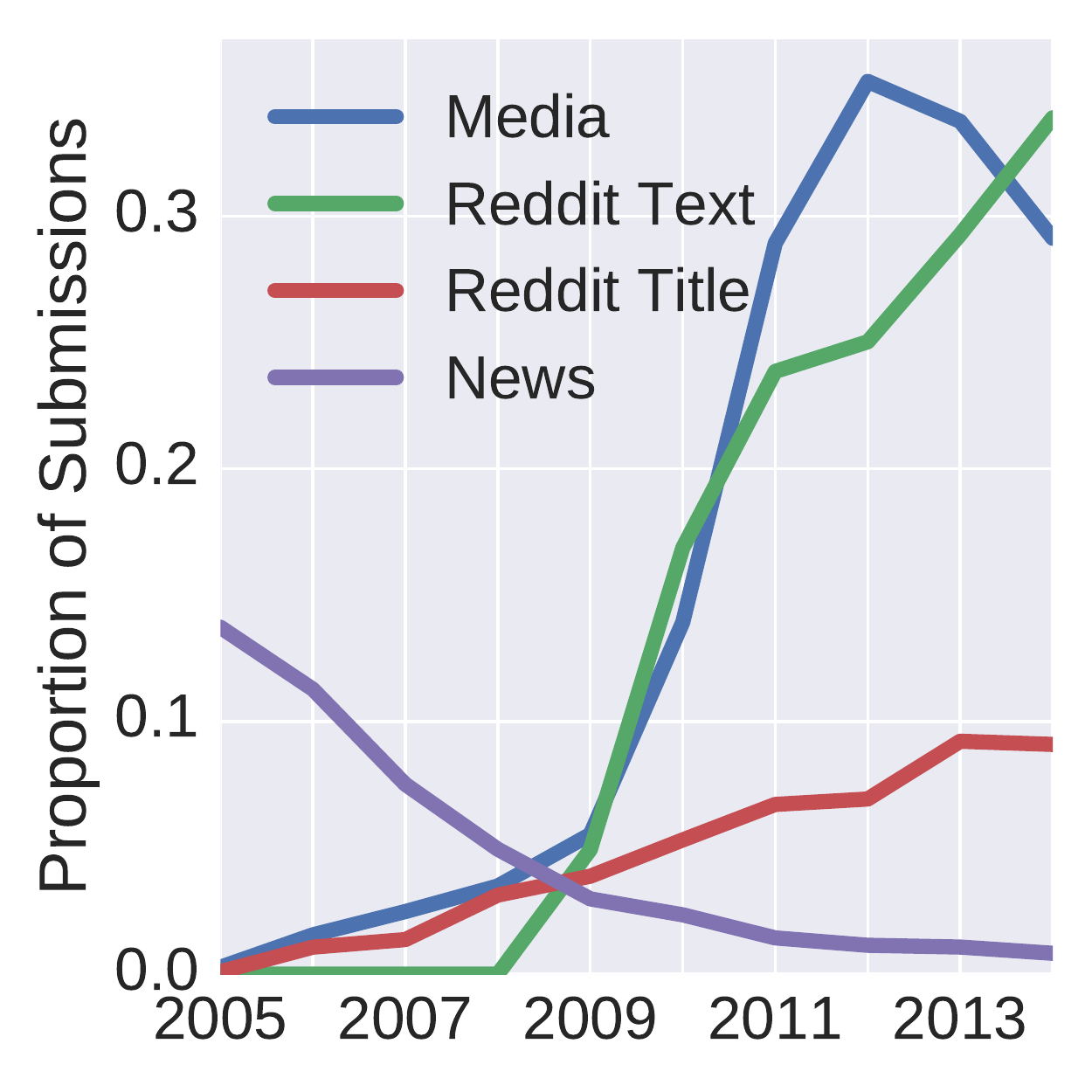}
  \captionof{figure}{Proportional popularity of types of Reddit posts
    over time across all subreddits.}
  \label{fig:domains}
\end{minipage}
\hspace{.001cm}
\begin{minipage}{.22\textwidth}
  \centering
  \includegraphics[width=\textwidth]{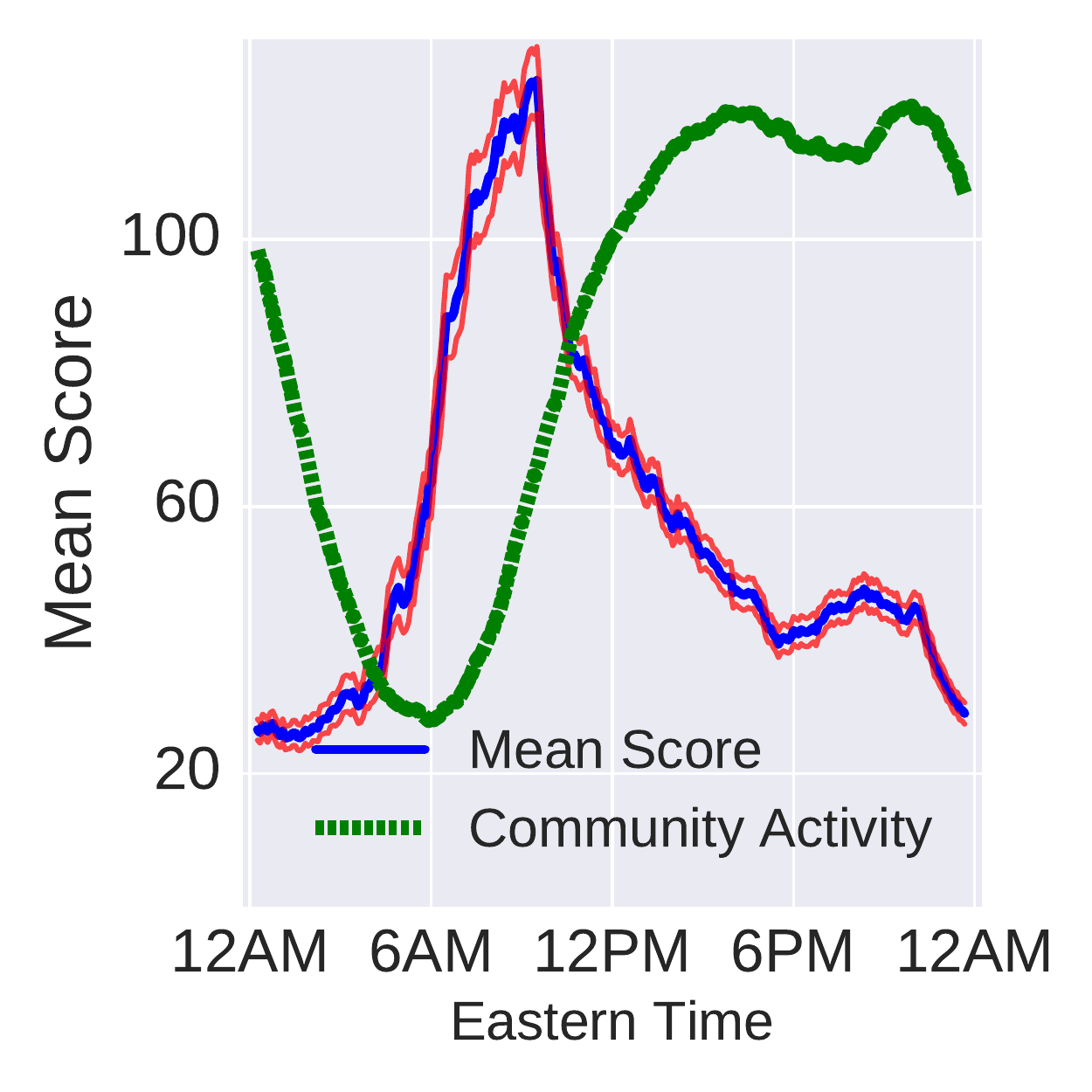}
  \captionof{figure}{Average score versus time of day (eastern) on
    \communityname{aww} with 95\% CI (red) and activity
    levels. }
  \label{fig:aww_plot}
\end{minipage}
\end{figure}

Many subreddits embody a larger growing trend towards images, video,
and other media content. Nearly all major social media sites (e.g.,
Facebook, Twitter, Pinterest) support image and video, and some
networks make multimodal content their focus (e.g., Instagram). We
performed a similar analysis to Singer et
al. \shortcite{singer2014evolution}, and hand-categorize popular
top-level domains on Reddit into ``media'' (e.g., \url{imgur},
\url{youtube}) ``news'' (e.g., \url{cnn}, \url{bbc}) and Reddit
internal title-only and text posts. Figure~\ref{fig:domains}
demonstrates the dramatic rise in multimedia content submitted to
Reddit from 2005 to 2014. Note that this graph is proportional --- the
raw number of multimedia submissions to the site is still rising, even
though the proportion has flattened. Roughly 30\% of all submissions
to Reddit are images, gifs, videos, and the like. In fact, more than
400 subreddits have each amassed more than 5,000 image submissions. If
researchers frame problems carefully, these communities offer a
diverse set of in-situ human and community reactions to multimedia
content without the need for expensive annotations.

We focus on six image-centric subreddits of varying popularity, visual
focus, and social structure. These communities range from
\communityname{pics},\footnote{According to the moderators:
  \communityname{pics} is ``a place to share photographs and
  pictures.''}  which has millions of subscribers and offers few
guidelines about what types of images are permitted, to
\communityname{RedditLaqueristas} [sic], where users submit
photographs of artistically lacquered fingernails. Typical examples of
image/text submissions made to \communityname{aww} are shown in in
Figure~\ref{fig:cats}.\footnote{The left submission was the more
  popular of the two, receiving at least 1K more upvotes than the
  right.}
General statistics about each of these datasets are presented in
Table~\ref{table:stats-pre}. Note that some community name
abbreviations are also introduced here.

While users are able to submit links from any website on the Internet
to any subreddit, the most common top-level domain is
\texttt{imgur.com} (\texttt{Alexa.com} rank 48, Fall 2016), a site
created to be an image hosting companion site to
Reddit.\footnote{\texttt{https://goo.gl/2fX34m}} Imgur allows users to
upload content which can subsequently be shared to Reddit. All images
in our datasets were fetched from Imgur.

\begin{table}
\centering
\begin{tabular}{lrrr}\\
 & \# Users & \#/\% Imgur & Cap Len\\
\midrule
\communityname{pics} & 2108K & 2472K/70\% & 9.84\\
\communityname{aww} & 1010K & 954K/81\% & 9.13\\
\communityname{cats} & 109K & 100K/73\% & 8.97\\
\communityname{MakeupAddiction (MA)} & 77K & 58K/57\% & 13.67\\
\communityname{FoodPorn (FP)} & 74K & 50K/77\% & 9.39\\
\communityname{RedditLaqueristas (RL)} & 27K & 39K/73\% & 11.12\\
\bottomrule\\
\end{tabular}
\caption{Number of unique users, number of Imgur submissions, and the
  average caption length for the communities used in this study. The
  number of unique users includes those who commented or submitted.}
\label{table:stats-pre}
\end{table}

We define a subreddit to be ``active'' if it receives more than 15
submissions on that day.\footnote{This is mostly done to filter out
  the unreliable feedback early in a community's life. After the first
  active day, the proportion of active days thereafter varies from
  96\% in the case of \communityname{pics} to 55\% in the case of
  \communityname{RL}, with an average of 83\% over all datasets.} We
attempted to scrape all images from all active days from all six
subreddits from Reddit's inception until February 1st, 2014.

As preprocessing we remove any duplicate images,
\footnote{We filter duplicate links by matching imgur ID and duplicate
  images by {\tt PHash} with a hand-picked hamming distance threshold
  of five. We attempt to discard \emph{all} copies of repeat
  submissions to mitigate any effect of repeated submissions, though
  deleted posts and pathological cases prevent us from guaranteeing
  that there are no duplicates.} and any animated or corrupted
images. Imgur albums consisting of multiple images are also
discarded. All images are resized to 256 pixels by 256 pixels. All
datasets, including train/test splits, are available at
\href{www.cs.cornell.edu/~jhessel/cats/cats.html}{\texttt{www.cs.cornell.edu/\textasciitilde{}jhessel/cats/cats.html}}.

\section{Time and Rich-getting-richer}
\label{sec:eda}

Our objective is to isolate content features and predict the relative
popularity of two items posted at approximately the same time. This
approach has the advantage that it is relatively insensitive to two
factors: the time of posting and the absolute number of positive user
votes. In this section we provide arguments that these factors are
significant in our data set and that previous methods for controlling
for these factors are not sufficient.

\paragraph{Why Control for Time?}
Raw post scores are influenced by timing factors in complex and
difficult-to-measure ways.  Reddit is a dynamic, evolving platform, so
expected popularity of submissions varies across many time scales.
In Figure ~\ref{fig:aww_plot}, for example, we show the mean score of
submissions made at various times during the day averaged over a
sliding 30 minute window in \communityname{aww}.  The figure also
shows the average activity level of the community as measured by
number of submissions. There is a dramatic spike in average submission
score for posts submitted at 9AM when compared to posts submitted at
6AM or 12PM.

\begin{figure}
  \centering
  \begin{subfigure}{.22\textwidth}
    \centering
    \includegraphics[width=\linewidth]{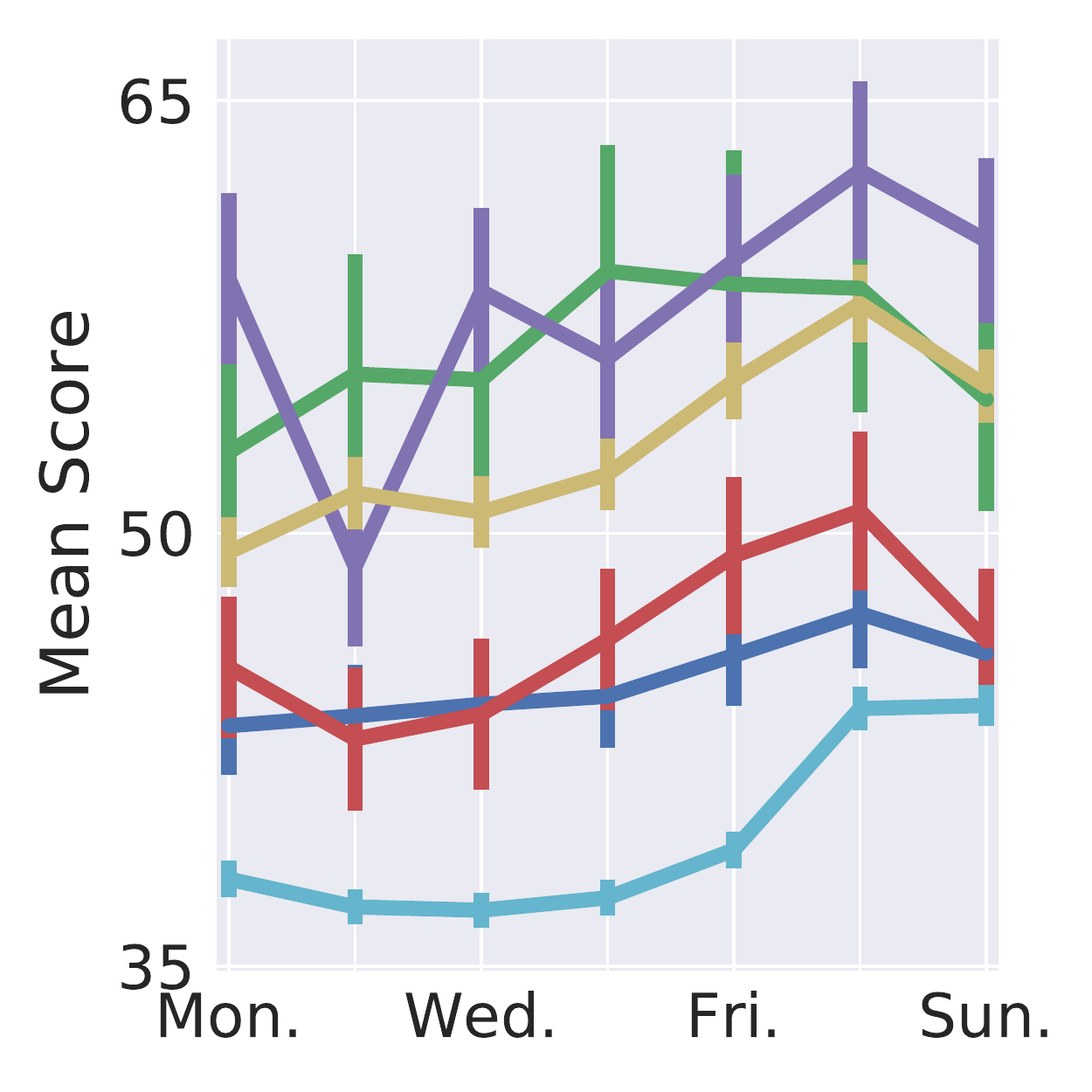}
    \caption{Day of week versus score  }
    \label{fig:day_plot}
  \end{subfigure}
  \hspace{.001cm}
  \begin{subfigure}{.22\textwidth}
    \centering
    \includegraphics[width=\linewidth]{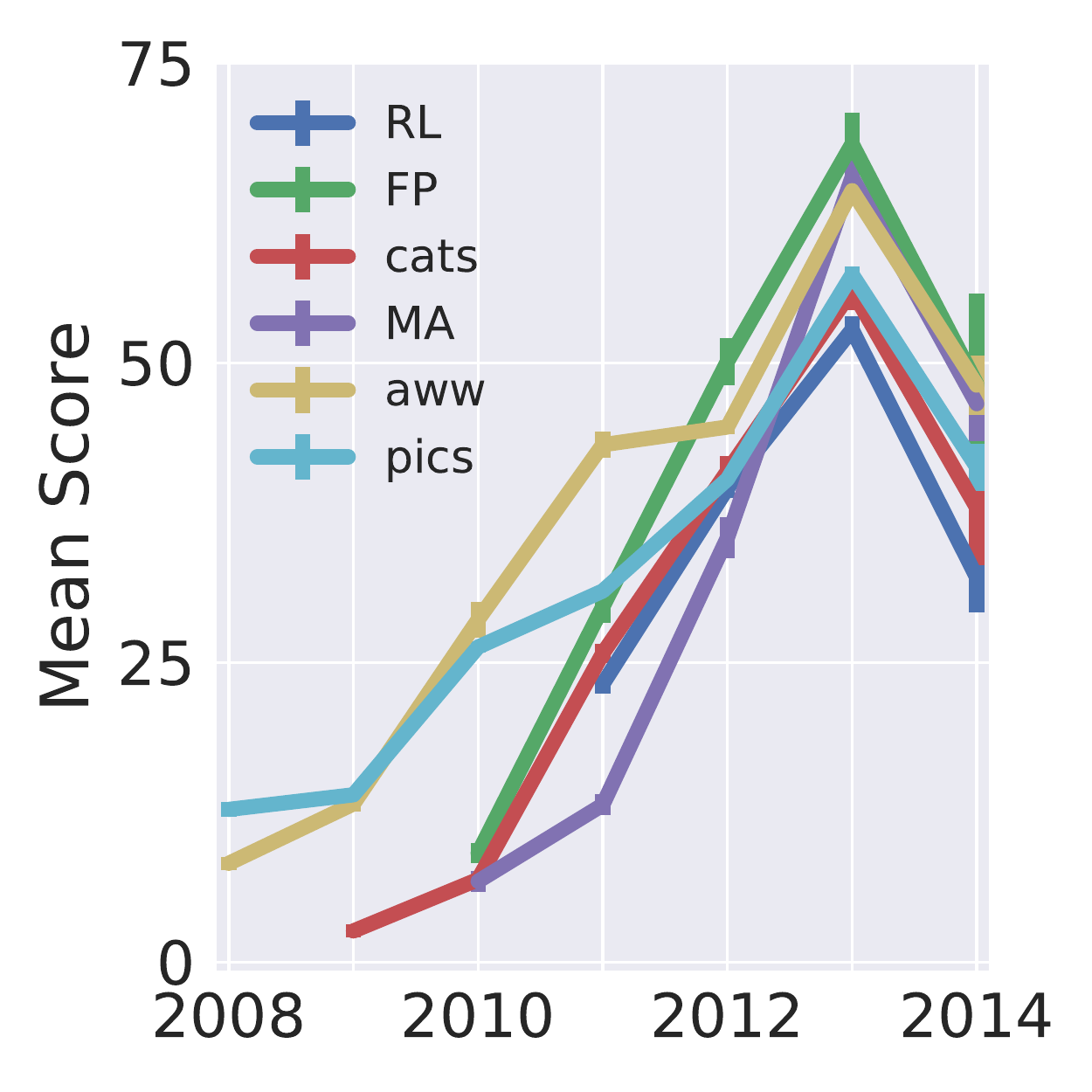}
    \caption{Year versus score  }
    \label{fig:year_plot}
  \end{subfigure}
\caption{Relationship between various measures of time and eventual
  submission score with 95\% confidence intervals.}
\label{fig:time_plots}
\end{figure}

Expected popularity also varies periodically between days of the
week. Figure~\ref{fig:day_plot} shows posts binned by day of the
week. The average score of submissions to \communityname{aww},
\communityname{pics}, and \communityname{cats} seems to be greater on
weekends when compared to weekdays.  These patterns are not always
easily modeled; the number of upvotes in
\communityname{MakeupAddiction} falls sharply on Tuesdays, potentially
as a result of the community's ``Text Tuesdays'' tradition (when only
text posts are allowed). We observed similar patterns in the other
subreddits. Figure~\ref{fig:year_plot} illustrates binning by
submission year. The average post score on Reddit seems to be
increasing over time, but it is unclear whether this trend has
continued in 2014, as vote totals might not have had the chance to
stabilize at the time of scraping in early 2014.

Reddit communities more closely resemble time-sensitive ``cultural
markets'' as described by Salganik et
al. \cite{salganik2006experimental} than any of the three
image-sharing settings described by Khosla et
al. \cite{khosla2014makes}.

\paragraph{Mean Normalization.}
Lakkaraju et al.'s \shortcite{lakkaraju2013s} work offers a starting
point for designing a time-control mechanism. Their original goal was
to control for popularity \emph{between} subreddits, whereas we aim to
control for time \emph{within} a community.

We identified several problems when applying their time control
method, which we call \emph{mean normalization} (MN), to our
setting. MN divides the score of a Reddit post by the average score of
all posts surrounding it in an hour. Estimating a robust and accurate
mean is difficult because of the dynamics of popularity. The
submission distribution is skewed by rich-get-richer processes (for
\communityname{aww}, average skew is 7.33, average kurtosis is 69.25),
so \emph{average} popularity as a statistic does not capture a fair
notion of quality. Furthermore, submissions that are unlucky enough to
posted within the same hour as a popular post are unfairly
downweighted by the rich-get-richer process.

For some subreddits, a one hour time window is probably too
big. Figure ~\ref{fig:aww_plot} suggests that one hour can encompass
large changes in expected popularity in the fast-paced world of
Reddit, e.g., the mean score for 6AM submissions is around $64$,
whereas the mean score for 7AM submissions is around $93$.

Finally, in less popular communities, Lakkaraju et al. note that it is
often difficult to get a stable estimate of the mean submission
popularity within an hour. While their setting doesn't require
estimating this mean, ours does. In \communityname{FoodPorn}, for
example, just 44\% of all submissions have at least 5 in-window
submissions ($\mu = 4.57$) to take the mean over.

\paragraph{Raw Transformations on Reddit.}
Raw post scores, even normalized for temporal effects, may be too
noisy to learn accurate models. Self-reinforcing ``rich get richer''
dynamics in online interfaces result in complex, non-linear
relationships between quality and popularity
\shortcite{salganik2006experimental}. Furthermore, recent work shows
that these dynamics differ significantly from community to community;
sometimes a small number of highly scored submissions is preferred,
while in other cases, the scores are more evenly distributed
\cite{lee2016beyond}.
The complexity of this relationship is compounded by website interface
changes, ranking algorithm modifications,\footnote{In fact, between
  the time of submission and publication, Reddit \emph{did} entirely
  change their method for computing post scores:
  \href{https://goo.gl/zHcKzL}{\texttt{goo.gl/zHcKzL}}} and
innumerable other subtle effects.

Transformations of raw votes are known to be more effective than
highly-skewed raw values. Khosla et al. \cite{khosla2014makes}, for
example, successfully use a log-transformation on Flickr view
counts. In the case of Reddit, however, heuristic transformations like
these enforce complex biases that are not consistent between different
subreddits. Also, it is not clear how to extend these to a
time-controlled setting, in general.

\paragraph{Our Approach: Pairwise Sampling.}
Because only relative judgments need to be made, the comparison of
submissions made in quick succession requires no assumptions about the
skewness of the score distribution. We do not need to compute a stable
estimate of average popularity, so sparse submission data can be
handled. No ad-hoc transformation of raw scores is required,
either. If the time difference between two posts is small we can train
models using the assumption that posts start on roughly equal
footing. We can then quantify the validity of those assumptions in
terms of timing and \fuser baselines, and directly compare cats and
captions to creators and the clock.

While it would be ideal to design a pairing process that would control
for other social effects, doing so would be substantially more
difficult than accounting for time.
For example, if we sample pairs of posts made by the same author in a
short time window, we would lose---at the very minimum---the 75\% of
submissions made to \communityname{pics} by users who have deleted
their accounts or who only submit a single time. Also, Reddit enforces
a one-post-per-several-minutes submission rate on a majority of
accounts, meaning our stringent time controls would need to be
relaxed. We leave sophisticated user-identity controlling sampling
procedures to future work, and focus on quantifying the performance of
\fuser features instead.

After scraping the images associated with each subreddit, our goal is
to pair submissions to minimize differences in timing. The pairing
process is controlled by several parameters.  For each community we
define a fixed, maximal allowed time-window so that pairs are not too
far apart.  We select pairs greedily to minimize this gap, so in
practice the average time difference is smaller than the maximum
window size.  To mitigate the effects of noise, we force the score
difference between members of a pair to be at least 20,\footnote{A
  majority of experiments were also conducted with the minimum
  difference parameter set to four; results were similar to those
  presented here.} and the eventually more popular submission must
also be at least twice as popular as the other. Additionally, we
ignore posts that received a score of less than two to avoid spam and
other very low-quality submissions that received no upvotes.

Table~\ref{tab:stats-post} shows the maximum and average window sizes,
along with the number of pairs that were sampled using a simple greedy
algorithm. For \communityname{aww} and \communityname{pics}, the most
popular communities we examine, sampled pairs are submitted 15 seconds
apart on average.

\begin{table}
\centering
\begin{tabular}{lrrr}\\
 & Max/Avg Win & Med/Avg Diff & \# Pairs\\
\midrule
\communityname{pics} & 30/15 sec & 117/478 & 44K\\
\communityname{aww} & 30/15 sec & 90/393 & 33K\\
\communityname{cats} & 15/7 min & 69/231 & 15K\\
\communityname{MA} & 60/24 min & 88/227 & 10K\\
\communityname{FP} & 120/53 min & 62/188 & 8K\\
\communityname{RL} & 30/14 min & 56/118 & 9K\\
\bottomrule\\
\end{tabular}
\caption{Statistics regarding the sampling used to generate ranking
  pairs. The maximum window is the maximum number of minutes that two
  submissions can be apart to be paired up, whereas the average window
  is the average time between all sampled pairs. The median and mean
  score differences between pairs is also given.}
\label{tab:stats-post}
\end{table}

\begin{table}
\centering
\begin{tabular}{llllllll}%
\toprule
 & \communityname{aww} & \communityname{pics} & \communityname{cats} & \communityname{MA} & \communityname{FP} & \communityname{RL}\\

\midrule
Humans & 60.0 & 63.6 & 59.6 & 62.2 & 72.7 & 67.2\\

\bottomrule
\end{tabular}
\caption{Human annotation accuracy results.}
\label{tab:human_res}
\end{table}

\paragraph{Human Validation.}
We first consider the validity of this new task by conducting a small
human study. Our goal with this study was to determine if the task of
predicting relative engagement was even possible using these datasets,
or whether there is no correlation between content and Reddit score.
We asked annotators to predict which among two time-controlled
submissions they thought would get more upvotes. For each of the six
datasets we showed the same 20 pairs to annotators.\footnote{Due to a
  sampling bug, \communityname{pics} pairs in the human experiments
  were sampled from 2009-2012 instead of 2009-2014.} In total, we were
able to gather 1400 human pairwise judgments. In addition, users were
given the option of describing ``why'' they made the choice they did.

Annotators used a variety of techniques to make their
decisions. Rationale ranged from basic aesthetic observations (``Much
better photo;'' submitted with a correct annotation. ``Better photo;''
correct annotation. ``homemade + steak + picture resolution (so
profesh);'' correct annotation.) to comments about how unique images
were (``Dude, it's a cat with a pencil;'' incorrect). Sometimes, the
authors disqualified submissions based on the associated text, rather
than on the images (``Less begging in the title;'' incorrect.). Many
annotators used their perception of the communities when making
judgements (``The Internet loves meat;'' correct. ``Easy. Desserts
always win;'' correct.). Sometimes the annotators wished they were
more familiar with the community, e.g., one user submitted an
incorrect annotation, noting that ``[they were not] sure whether
FoodPorn is about the images or the food concept.'' Some pairs were
universally difficult. For example, 83\% of annotators incorrectly
selected a cute rabbit (``Dat bunny face;'' +10 Reddit score) over an
out-of-focus photo of a duck\footnote{One redditor comments regarding
  the misfocused image: ``That trashcan [in the background] is in
  excellent quality.''} with the caption ``My brother got a duck
yesterday..'' [sic] (+115 Reddit score).

The resulting mean accuracy for each dataset is presented in
Table~\ref{tab:human_res}. In general, humans are able to guess
pairwise rankings of submissions from images and captions, but the
task is difficult.\footnote{Because the human study only considered a
  small subset of image pairs, the exact values reported are less
  precise than for the other results: the 95\% confidence intervals
  for the human annotations are on average $\pm 6$} Having validated
that the task is neither trivial nor impossible for humans, we now
move on to our machine learning experiments.

\section{Model Design}

For relative popularity prediction, we use a pairwise learning-to-rank
model
\cite{herbrich1999large,joachims2002optimizing,burges2005learning}.
Specifically, our data is of the form $\{x_{1i}, x_{2i},
y_i\}_{i=1}^n$ where $\langle x_{1i}, x_{2i} \rangle$ is a pair of
Reddit submissions posted at similar times, and $y_i$ is an indicator
variable that encodes which submission became more popular. We train a
linear classifier on top of the \emph{vector difference} of two
entities for predicting which of the two is more highly ranked
(i.e. $y_i$). As such, we experiment with models of the form
\begin{equation}
\hat y_i = w^T(f(x_{1i}) - f(x_{2i}))
\end{equation}
where $w$ is a set of regression weights and $f$ is one of a variety
of Reddit submission representation functions. In all experiments, we
use a hinge loss, which is minimized with respect to the coefficients
of the regression itself and, if applicable, with respect to the
trainable parameters of $f$.

Note that our model implicitly learns a scoring function that can
assign a quality score to \emph{unpaired} examples. Specifically
$w^Tf(x) \in \mathbb{R}$ is a value that correlates with the model's
ranking of that submission.\footnote{Rather than approximating the
  global raw Reddit score ranking, the model \emph{induces} a ranking
  with desirable properties, e.g., it cannot be predicted from timing
  features.} This function could be used by moderators to compute
model scores of novel, incoming submissions. We use this function in a
later section to interpret our results.

\subsection*{Cats and Captions}
The textual and visual characteristics of the six communities we
examine are complex and varied. For example, most images in
\communityname{RL} are of fingernails, which are out of domain for
pretrained computer vision models. Similarly, complex social patterns
and tags emerge within language e.g., ``CCW'' meaning ``constructive
criticism welcome.'' As a result, a dataset-by-dataset examination of
specific, higher-level processes like image
\cite{khosla2012memorability} or text \cite{danescu2012you}
memorability transfer-learned from other domains is reserved for
future work. The goal of this section is \emph{not} to argue that
these models are the best. Rather, we will use these generic feature
extractors to demonstrate the importance of modeling content at all.

\paragraph{Image Models.}
We experiment with a combination of lower-level features and deep
neural network models to represent image content. This mix of models
is similar to those explored by Khosla et
al. \shortcite{khosla2014makes}.

The most basic building blocks of the human visual system are edges
and colors, and the presentation of these features might effect how
appealing an image is.
Previous work (e.g., \cite{bakhshi2015red}) has found that colors can
play a role in human response to visual content. As such, we examine a
set of color features \textbf{Color}, consisting of an $l_1$
normalized vector based on the RGB values of the colors in the
image. We use Khan et al.'s \shortcite{khan2013discriminative} 50
universal color descriptors
and extraction code to compute this vector for each image.

The second feature set is histogram of oriented gradients
\cite{dalal2005histograms}. \textbf{HOG} features capture localized
pixel gradients in an image. We use the HOG feature extractor in
OpenCV \cite{opencv_library} with default parameters and use random
projection to reduce the dimension of the resulting features to 2K
from 34K.

Next, we examine the \textbf{GIST} image descriptors
\cite{oliva2001modeling}, which aim to capture ``perceptual dimensions
(naturalness, openness, roughness, expansion, ruggedness) that
represent the dominant spatial structure of a scene.'' We use the
\texttt{pyleargist}\footnote{\href{https://bitbucket.org/ogrisel/pyleargist/}{\texttt{bitbucket.org/ogrisel/pyleargist/}}}
  library to extract these 960 features.

Recently, convolutional neural networks have been used to extract
high-level concepts from image data. We use the popular
\textbf{VGG-19} \cite{Simonyan14c} and \textbf{ResNet50}
\cite{he2015deep}. Both of these networks are used as feature
extractors\footnote{We found deeper residual networks and network
  fine-tuning to be unhelpful in early testing.} by taking the
final-layer activations from a set of weights trained for the ImageNet
\cite{deng2009imagenet} ILSVRC-2014 classification task. Building a
linear model over extracted features in this manner is known to offer
an ``astounding'' baseline \cite{sharif2014cnn}.

\paragraph{Text Models.}
We first examine a set of \textbf{Structural} features of
language. These include the message length (in tokens and characters),
the token-to-type ratio, and a ``punctuation proportion'' feature to
capture some signal about an author's use of non-alpha-numeric
characters (e.g., emoji).

We consider three models that do not use word order.
The bag-of-words assumption is valuable both because of its relative
simplicity and because of its high performance (see Hill et
al. \shortcite{hill2016learning} for some benchmarks). 
First we define a set of \textbf{Unigram} features by mapping each caption to
a vector of binary indicator variables.
Second we extract topic distributions using a specialized biterm topic
model \cite{yan2013biterm} designed for short texts. We use 20 topics
in all cases, and extract the resulting $l_1$ normalized
\textbf{Topic} distributions.  Third we use a variant\footnote{We
  apply $l_2$ normalization after the averaging step, and don't apply
  word-level dropout.} of the deep averaging network (\textbf{DAN})
\cite{iyyer2015deep}. This model averages a set of word embeddings and
feeds the result though a simple multilayer perceptron. We consider a
3-layer DAN with 128 hidden units. The model's word embeddings are
tuned from a 100D GloVe \cite{pennington2014glove} pretrained set.
We also consider sequence-based features, specifically an
order-sensitive recurrent neural network. We train an \textbf{LSTM}
\cite{hochreiter1997long} on the sequence of words in a caption. The
parameters of the RNN are learned, and the word embeddings are tuned
from the same 100-dimensional starting vectors as the DAN. For
completeness, we also consider a bidirectional LSTM, \textbf{Bi-LSTM}
\cite{graves2005framewise}.

\subsection*{Creators and the Clock}

\begin{figure}
  \centering
  \begin{subfigure}{.23\textwidth}
    \centering
    \includegraphics[width=\linewidth]{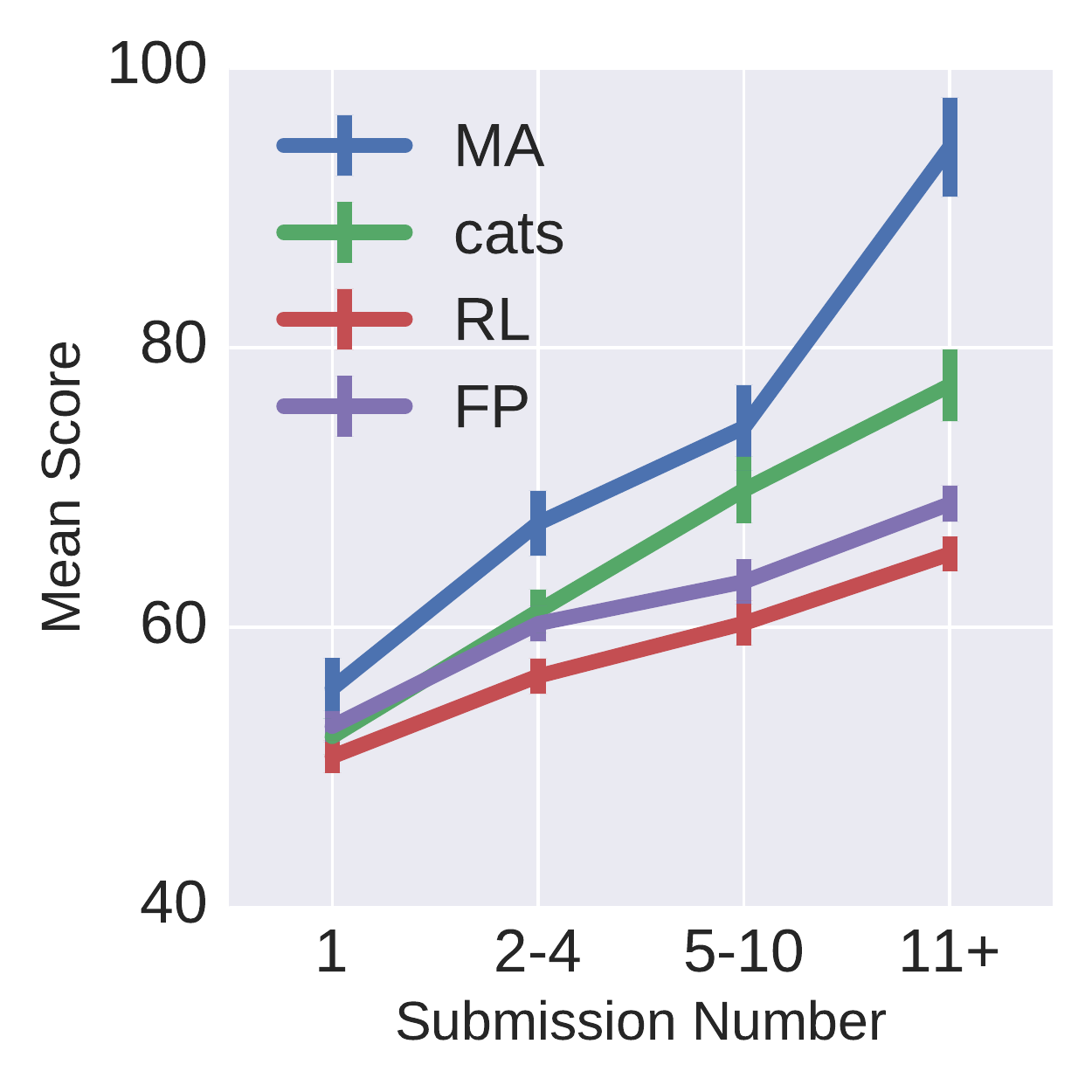}
    \captionof{figure}{Submission \# vs score}
    \label{fig:submission_plot}
  \end{subfigure}
  \begin{subfigure}{.23\textwidth}
    \centering
    \includegraphics[width=\linewidth]{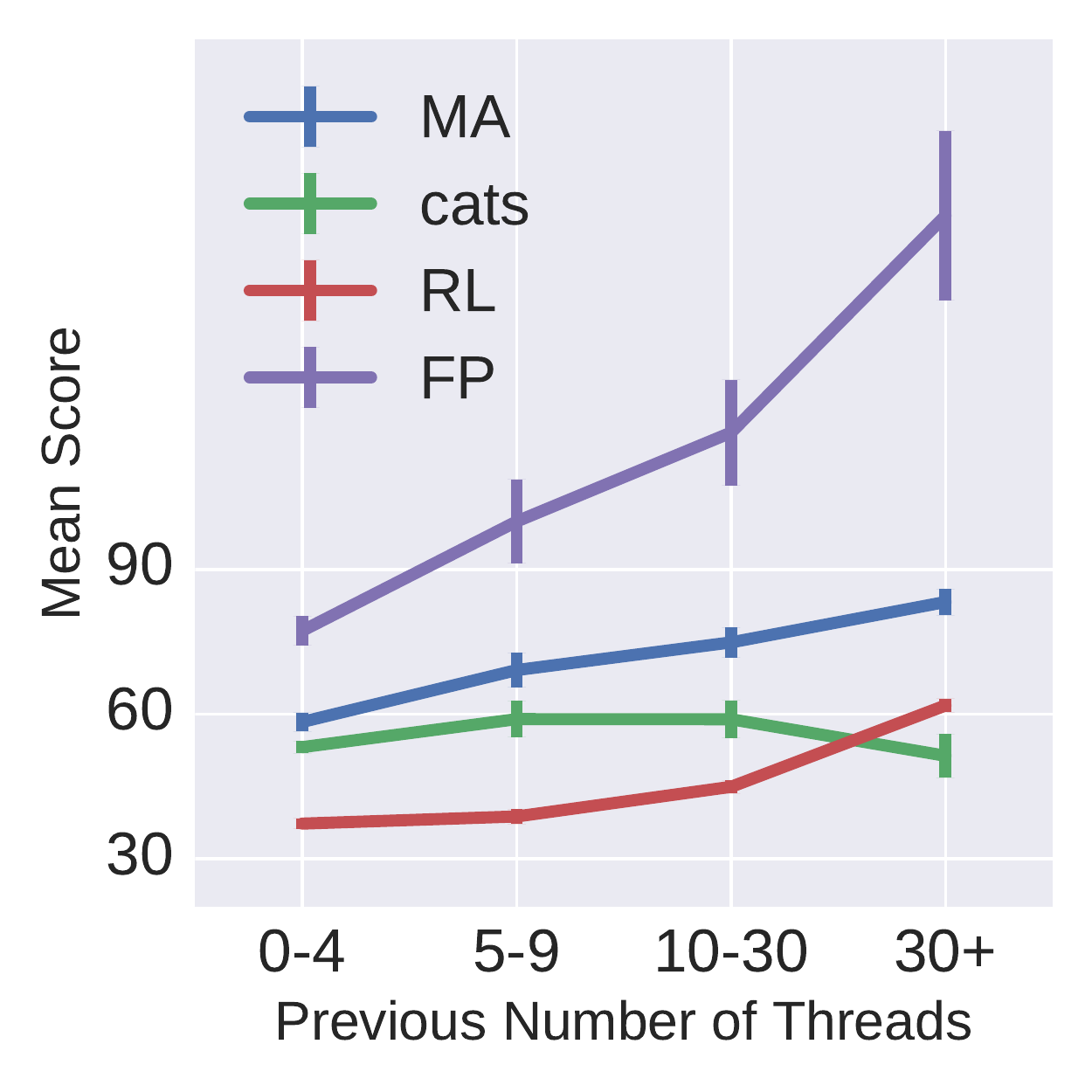}
    \captionof{figure}{Conversation \# vs score}
    \label{fig:conversation_plot}
  \end{subfigure}
\caption{Relationship between various measures of time and eventual
  submission score for several subreddits, with 95\% CI.}
\label{fig:pre_interact_plots}
\end{figure}

\paragraph{\fUser Features.}
Can Reddit users get upvotes based on an attained status as on other
social networks? An explicit and persistent user identity exists for
some users on \communityname{MakeupAddiction} and
\communityname{RedditLaqueristas} in the form of a \emph{flair} that
is displayed alongside a given user's posts. Most often, the flair
contains a link to a given user's Instagram profile.

For other communities, however, a majority of users submit only a few
times. Around 60\% of submissions made to \communityname{aww} and
\communityname{cats}, for example, are made by users who submit at
most three times ever. Even if a celebrity status were earned and
upvote counts were artificially inflated as a result, in these
communities, this likely plays a lesser role.

Another hypothesis is that as a user gains familiarity with a
community, they are better able to submit content of interest to that
community. Indeed, a user who has a better sense of the types of content
popular in a community might be more likely to submit high quality
content than a newcomer.

Even though we cannot disentangle the effects of status and
experience, we can still define features that capture aspects of a submitter's previous
behavior within a community.
Such features have previously been used in studies on Reddit
\cite{althoff2014ask,jaech2015talking} and Slashdot
\cite{lampe2004slash}, among others.

Two easily measured quantities are how many times a user has
previously submitted and how many total threads a user has previously
interacted with. Figure~\ref{fig:submission_plot} and
Figure~\ref{fig:conversation_plot} show that correlations exist
between score and previous interactions. In
\communityname{redditLaqueristas}, for example, if a submission is a
user's fifth to tenth, it is more likely to receive upvotes than if a
submission is a user's first. In \communityname{cats}, however,
participating in more than 30 threads by commenting or posting seems
to be associated with slightly lower average popularity.

The following set of \fuser features are computed for each submitter
at the time of their submission. When a statistic is not properly
defined for a given user at a given time (e.g., average previous
comment length when they have no previous comments; the submitter
deleted their account, etc.) the mean value over the training set is
substituted.\footnote{We ran \fuser experiments considering only pairs
  that consisted of no deleted users and no users without previous
  interaction data; the results were comparable.}

Previous work (e.g., \cite{dror2012churn}) has found information
regarding \emph{how much} a user participates in a community to be a
useful predictor of their future behavior. The \textbf{Activity}
feature set includes the number of previous posts/comments, how long
the user has been a member of the community, the time since previous
interaction, and the ratio of posts to posts plus comments for that
author.

It is possible that \emph{how} a user interacts with others in a
community is more important than how much they interact with a
community.

The \textbf{Type} feature set includes average comment
length, average comment token-to-type ratio, average conversation tree
depth of comment, the proportion of previous comments with replies,
the proportion of previous submissions wherein the user commented
multiple times, and median time-to-response from thread start.

Several variables are used to quantify the community-perceived
\textbf{Quality} of a submitter's previous interactions. Instead of
using statistics based on raw scores, which can be skewed by a
small number of very popular interactions, we use Jaech et al.'s
\shortcite{jaech2015talking} $k$-index, which counts the number of times
a user has submitted either a post or a comment that received more
than $k$ upvotes. To normalize for a user's total activity, we divide by
the total number of posts/comments that user made to form a statistic
we call $k$-rate. We compute $k$-rate for $k \in \{5, 10, 50, 100\}$ for
both posts and comments. While the quality statistics might leak
timing information, we would like the \fuser baseline to be as strong as
possible.

\paragraph{Timing Models.}
In the pairwise ranking setting, a \textbf{Random} guess is correct
half of the time. Furthermore, it is possible that the post that was
created \textbf{Earlier} has a tendency to get more upvotes because it
has existed longer, so choosing the submission in the pair that was
posted first makes for a good baseline.

Finally, we include a \textbf{Time} baseline to quantify how well the
pairing process controls for time. Instead of attempting to
hand-design a set of rules, because the effects of time are
complicated we choose to simply learn a time feature classifier. We
use a 1-hot encoding of the minute-in-hour, hour-in-day, day-in-week,
and year of the post. Instead of subtracting the resulting encodings,
we concatenate to give the model access to the absolute and relative
timing. The classifier we use is a one-hidden-layer neural network
with 100 hidden units to capture potential non-linear relationships.

\section{Results}

For all experiments, we compute 15-fold cross validation accuracy in an
80/20 train/test split. We withhold 10\% of training data as a
validation set, which is used to optimize regularization parameters
and for early stopping. Models are trained using Keras
\cite{chollet2015keras} with the Theano \cite{2016arXiv160502688short}
backend.

\subsection{Unimodal Experiments}

\begin{table}[t]
\centering
\begin{tabular}{l|lllllll}%
\toprule
 & & \communityname{aww} & \communityname{pics} & \communityname{cats} & \communityname{MA} & \communityname{FP} & \communityname{RL}\\

\midrule
\multirow{3}{*}{\rotatebox[origin=c]{90}{\parbox[c]{1cm}{\centering Timing}}}  & Random & 50.0 & 50.0 & 50.0 & 50.0 & 50.0 & 50.0\\
 & Earlier & \emph{51.7} & \emph{51.1} & \emph{49.9} & 48.9 & \emph{48.6} & 48.7\\
 & Time & 50.2 & 50.2 & \emph{50.7} & \emph{50.4} & \emph{49.7} & \emph{50.6}\\

\midrule
\multirow{3}{*}{\rotatebox[origin=c]{90}{\parbox[c]{1cm}{\centering User}}}  & Type & 50.6 & 51.2 & 50.7 & 52.8 & 51.8 & 56.1\\
 & Activity & 51.1 & 53.6 & \emph{52.8} & 55.0 & 53.9 & 60.6\\
 & Quality & \emph{54.7} & \emph{55.5} & \emph{52.9} & \emph{60.7} & \emph{55.5} & \textbf{\underline{67.3}}\\

\midrule \midrule
\multirow{6}{*}{\rotatebox[origin=c]{90}{\parbox[c]{1cm}{\centering Textual}}}  & Struct & 56.2 & 54.8 & 56.5 & 50.9 & 52.3 & 52.5\\
 & Topic & 55.2 & 55.8 & 56.8 & 60.4 & 55.2 & 55.5\\
 & DAN & 58.6 & \emph{58.3} & 58.5 & \emph{62.2} & \emph{57.6} & 59.8\\
 & LSTM & \emph{59.4} & \emph{58.8} & \emph{58.7} & 61.0 & \emph{57.0} & 59.1\\
 & Bi-LSTM & \emph{59.7} & \emph{58.9} & \emph{59.3} & \emph{61.8} & \emph{57.8} & \emph{59.6}\\
 & Unigram & \emph{59.7} & \emph{58.6} & \emph{59.5} & \emph{63.0} & \emph{57.6} & \emph{60.8}\\

\midrule
\multirow{5}{*}{\rotatebox[origin=c]{90}{\parbox[c]{1cm}{\centering Visual}}}  & HOG & 51.7 & 52.8 & 51.9 & 53.5 & 53.5 & 53.5\\
 & GIST & 52.7 & 53.0 & 53.5 & 55.9 & 56.5 & 56.3\\
 & ColorHist & 55.3 & 53.7 & 55.6 & 55.0 & 56.5 & 54.5\\
 & VGG-19 & 63.4 & 58.9 & 61.1 & 62.4 & 62.8 & 62.1\\
 & ResNet50 & \textbf{\underline{64.8}} & \textbf{\underline{60.0}} & \textbf{\underline{62.6}} & \textbf{\underline{64.9}} & \textbf{\underline{65.2}} & \emph{64.2}\\

\bottomrule
\end{tabular}
\caption{Unimodal accuracy results averaged over 15 cross-validation
  splits; higher accuracy is better. Bolded results are the best in
  the whole column and are underlined if differences are
  significant. Italicized results are tied for the best among their
  feature type. 95\% CI are on average $\pm .5$ and never exceed $\pm
  1$ for the non-timing features.}
\label{tab:unimodal_res}
\end{table}

Next we assess the individual ability of each modality to predict the
eventual popularity of content. The results for each
dataset and feature set are given in Table~\ref{tab:unimodal_res}.
Because the classification problem is a balanced two class task, we
only report accuracy.

\paragraph{Pairwise Ranking Controls for Time.}
Our objective in using pairwise ranking is to reduce the effect of
time-of-posting as a confounding factor.  As shown in Section
\ref{sec:eda}, time-based features are, in general, strongly
predictive of average user engagement.  But in the pairwise ranking
setting, we were happy to see that neither the learned time classifier
nor the ``earlier'' baseline were able to achieve meaningful
performance above random. This suggests that we are effectively
controlling for time of posting.

\paragraph{Previous Quality Predicts Current Quality.}
Among \fuser features, quality of previous submitted content is the
best predictor of future success. The particular types of interactions
(e.g., posts vs. comments, comment length) also seem to be less
important than the absolute volume of previous interactions.

\paragraph{For Words, Simpler is Better.}
Order-sensitive and deeper models models rarely outperformed the
shallower, order-unaware unigram models. Interestingly, structural
features performed particularly well on \communityname{cats} and
\communityname{aww}; we observed that longer, story-like titles worked
well in both of those communities. For all datasets, the best
text-only models performed worse than the best image-only models,
suggesting that visual content is more predictive of relative
popularity than textual content in these communities.

\paragraph{For Images, More Complicated is Better.}
For all datasets, the best performing image algorithm was the deep
neural network ResNet50.  The fact that ResNet50 outperformed its
shallower counterpart VGG-19 suggests that this task is well-formulated
as a computer vision task. In general, the CNN approaches performed
better than the lower-level image features, though all outperformed
random.

\subsection{Multimodal Experiments}

\begin{table}[t]
\centering
\begin{tabular}{llllllll}%
\toprule
 & \communityname{aww} & \communityname{pics} & \communityname{cats} & \communityname{MA} & \communityname{FP} & \communityname{RL}\\

\midrule
Time \texttt{+} User & 54.1 & 54.7 & 52.1 & 58.8 & 54.2 & 64.8\\
All User & 56.3 & 55.3 & 54.6 & 60.9 & 56.0 & \textbf{\underline{68.4}}\\
ResNet50 & 64.8 & 60.0 & 62.6 & 64.9 & \emph{65.2} & 64.2\\
Text \texttt{+} Image & \textbf{\underline{67.1}} & \textbf{\underline{62.7}} & \textbf{\underline{65.9}} & \textbf{\underline{67.7}} & \textbf{65.8} & 66.4\\

\bottomrule
\end{tabular}
\caption{Multimodal accuracy results averaged over 15 cross-validation
  splits. Higher accuracy is better, and accenting follows
  Table~\ref{tab:unimodal_res}. 95\% CI are on average $\pm .5$ and
  never exceed $\pm .76$. The best unimodal model ResNet50 is
  generally outperformed by the multimodal model, Text \texttt{+}
  Image. \fUser features alone (All \fUser) generally perform better
  on their own than when they are combined with timing features.}
\label{tab:multimodal_res}
\end{table}

We now directly compare ``cats and captions'' to ``creators and the
clock.'' In particular, given the high performance of unigram and
ResNet50 features, we use Lynch et al.'s \cite{lynch2016images}
elastic net regression method to jointly represent visual and textual
content, and call the model \textbf{Text \texttt{+} Image}. Because
timing features weren't found to be helpful when concatenated with
\fuser features (\textbf{Time \texttt{+} User}), we also include a
concatenation of all \fuser features, \textbf{All \fUser}. These
results are presented in Table~\ref{tab:multimodal_res}.

In five of six cases, content features outperform the \fuser features
for relative popularity prediction. In terms of relative improvement
over random, the magnitude of this improvement is between 245\% for
\communityname{cats} and 62\% for \communityname{MakeupAddiction}.  In
five of six cases performance significantly improves when we combine
text and images, indicating that this task is well-formulated as a
multimodal task. In these cases, the relative improvement over random
when adding text to the best image model varies between 27\% for
\communityname{pics} and 16\% for \communityname{aww}.

\begin{table}[t]
\centering
\begin{tabular}{llllllll}%
\toprule
 & \communityname{aww} & \communityname{pics} & \communityname{cats} & \communityname{MA} & \communityname{FP} & \communityname{RL}\\

\midrule
Time \texttt{+} User & 55.5 & 51.7 & 52.6 & 56.9 & 52.8 & 60.5\\
All User & 60.4 & 51.0 & 54.3 & \textbf{63.1} & 57.9 & \textbf{66.0}\\
Text \texttt{+} Image & \textbf{65.5} & \textbf{66.0} & \textbf{67.3} & 62.7 & \textbf{62.6} & 65.4\\

\midrule
\bottomrule
\end{tabular}
\caption{Heldout, out-of-domain task accuracy results; bolded are
  best.}
\label{tab:holdout_res}
\end{table}

\paragraph{Fully-held Out, Different Distribution Test.}
One useful property of the models we consider is the unpaired scoring
function implicitly learned in the ranking process. While this scoring
function could be used to process novel submissions made to a
community, it's unclear how well patterns learned across training data
would generalize to testing data. Changing linguistic
\cite{danescu2013no} and visual \cite{wu2016time} preferences of
communities complicate this task considerably.

We selected 1000 pairs from each community sampled outside of the
training data's time span, and therefore out of the exact distribution
of the training data. These pairs were \emph{fully held out} meaning
that we evaluated them \emph{exactly once} for each model. The
accuracy of the content model and the \fuser/timing model in the
fully-held-out settings are given in Table~\ref{tab:holdout_res}.

While it is difficult to extrapolate from point estimates, the
fully-held out results display interesting changes in performance. In
particular, while differences in performance are relatively minor
(indicating that we likely didn't overfit) we see a roughly 28\%
decrease in performance in \communityname{MakeupAddiction}.  We find
some evidence suggesting that the community has evolved during the 10
month heldout period. In particular, for the image + text models, the
average posting time of the correctly-classified pairs is 11 days
earlier (and closer in time to the training data) than the average
posting time of incorrectly-classified pairs. Because only 1K held-out
pairs are considered, the statistical significance of this potential
difference cannot be established for \emph{all} models. However, this
pattern was observed across several models we considered.
Collectively, these observations suggest a potentially complex
relationship between training set generalizability and time.

\paragraph{Model Score vs. Raw Score.} Using traditional ranking
metrics in this pairwise setting is difficult because, as we have
argued above, there is no appropriate ``gold standard'' ranking to
compare against.  The scores received on Reddit would indeed provide
{\em a} ranking, but not an {\em appropriate} ranking, because those
scores are biased by precisely factors like timing we have discussed
and constructed our pairwise task to mitigate.
As a result, applying evaluation metrics like mean reciprocal rank
(MRR) or precision-at-K (p@K) that assume a ground-truth ranking is
not possible.

\begin{figure}
  \centering
  \includegraphics[width=.7\linewidth]{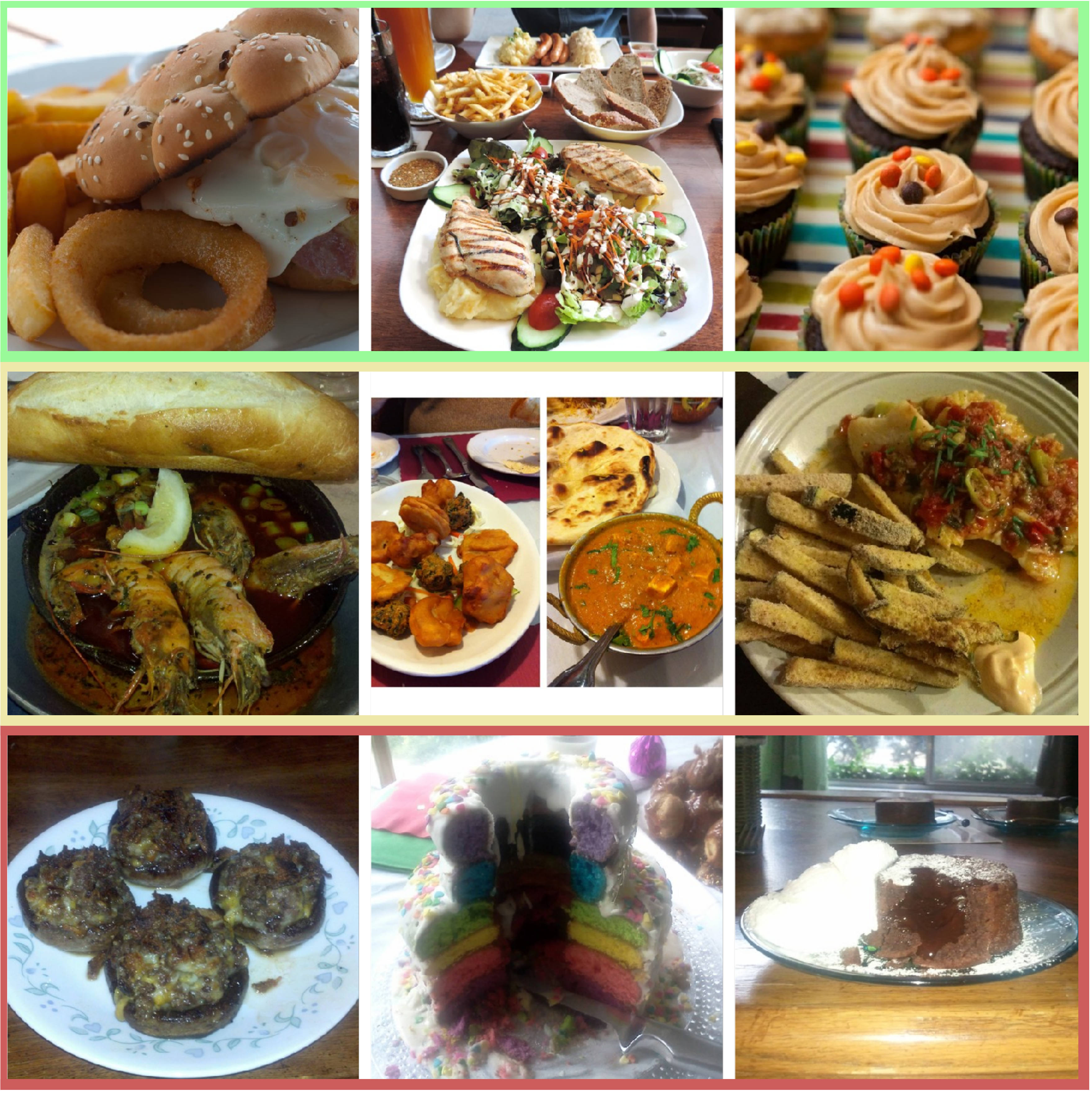}
  \caption{Examples from \communityname{FoodPorn} automatically scored
    by the ResNet50 model. The top, middle, and bottom rows are
    sampled from the 99th, 50th, and 1st percentiles of model scores
    respectively. While lighting effects likely relate to model
    scores, the underperformance of the color-only classifier and the
    performance jump when switching from VGG-19 to ResNet50 suggest
    that this is a rich computer vision task. Images courtesy
    \texttt{imgur.com}.}
  \label{fig:FoodPorn}
\end{figure}%

\newcommand{\awwtab}{\rowcolor{lightgreen}Found this 3 day old baby under a car in...\\
\rowcolor{lightgreen}This is Dexter. A year 1/2 ago my friend...\\
\rowcolor{lightgreen}My very first dog and best friend is...\\
\rowcolor{lightgreen}This sweet girl was found on the side of...\\
\rowcolor{lightgreen}Found this little guy getting rocks...\\
\rowcolor{lightyellow}Every time we start the boat...\\
\rowcolor{lightyellow}Henry is quite the lady killer\\
\rowcolor{lightyellow}Reddit, meet Sutton! My new kitty baby!\\
\rowcolor{lightyellow}Cloudy forgets to close his mouth...\\
\rowcolor{lightyellow}First post! My begging English Mastiff.\\
\rowcolor{lightred}Soft kitty, warm kitty, little ball of...\\
\rowcolor{lightred}My sleepy kitty enjoying the sun\\
\rowcolor{lightred}Happy kitty, sleepy kitty and man she...\\
\rowcolor{lightred}Shih Tzu + Beagle = Adorable\\
\rowcolor{lightred}GO! GO! GO! GO! GO! GO! GO!...\\
}
\newcommand{\awwmm}{\rowcolor{lightgreen} \raisebox{-.85\height}{\includegraphics[width=.32\textwidth]{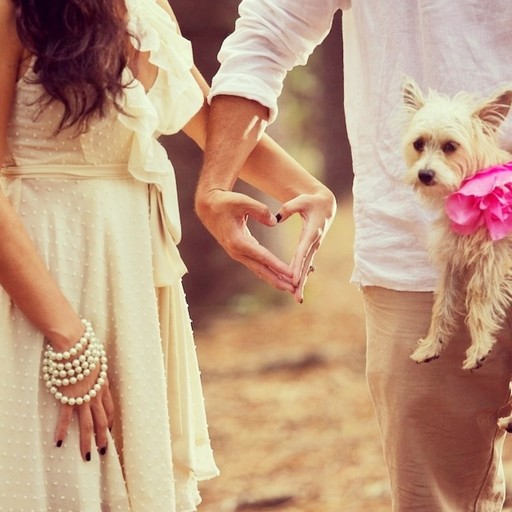}} & {\small We were taking a family photo but our dog kept scratching on our legs trying to be in our photo. We...}\\
\rowcolor{lightyellow} \raisebox{-.85\height}{\includegraphics[width=.32\textwidth]{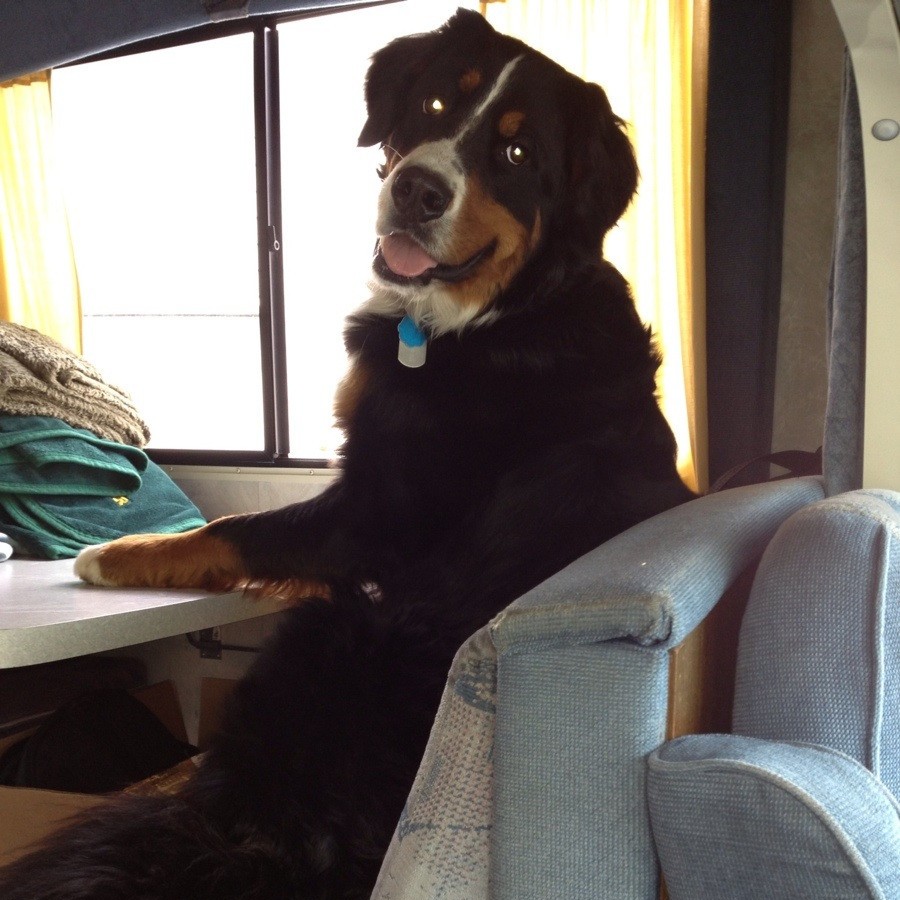}} & {\small Walter is ready for dinner.}\\
\rowcolor{lightred} \raisebox{-.85\height}{\includegraphics[width=.32\textwidth]{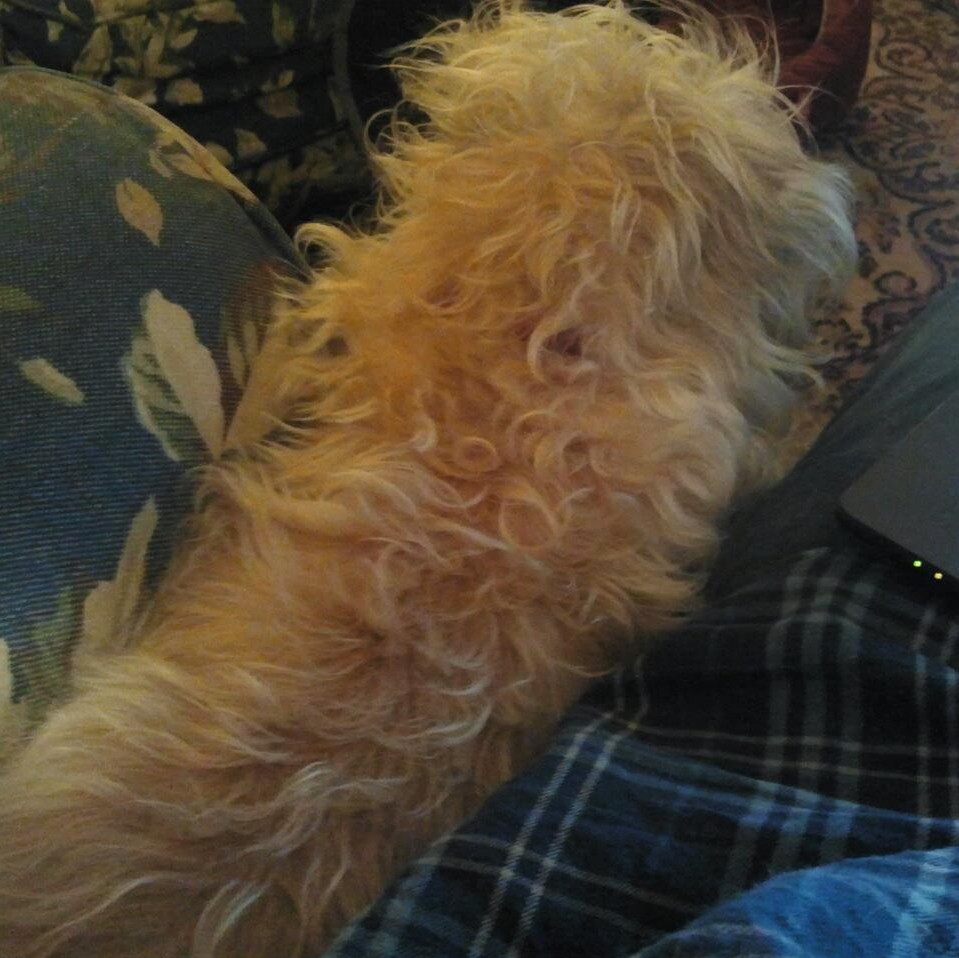}} & {\small When Hamish fits, he sits.}\\
}
\begin{figure*}[h]
\begin{subfigure}{.33\textwidth}
  \centering
  \includegraphics[width=.96\linewidth]{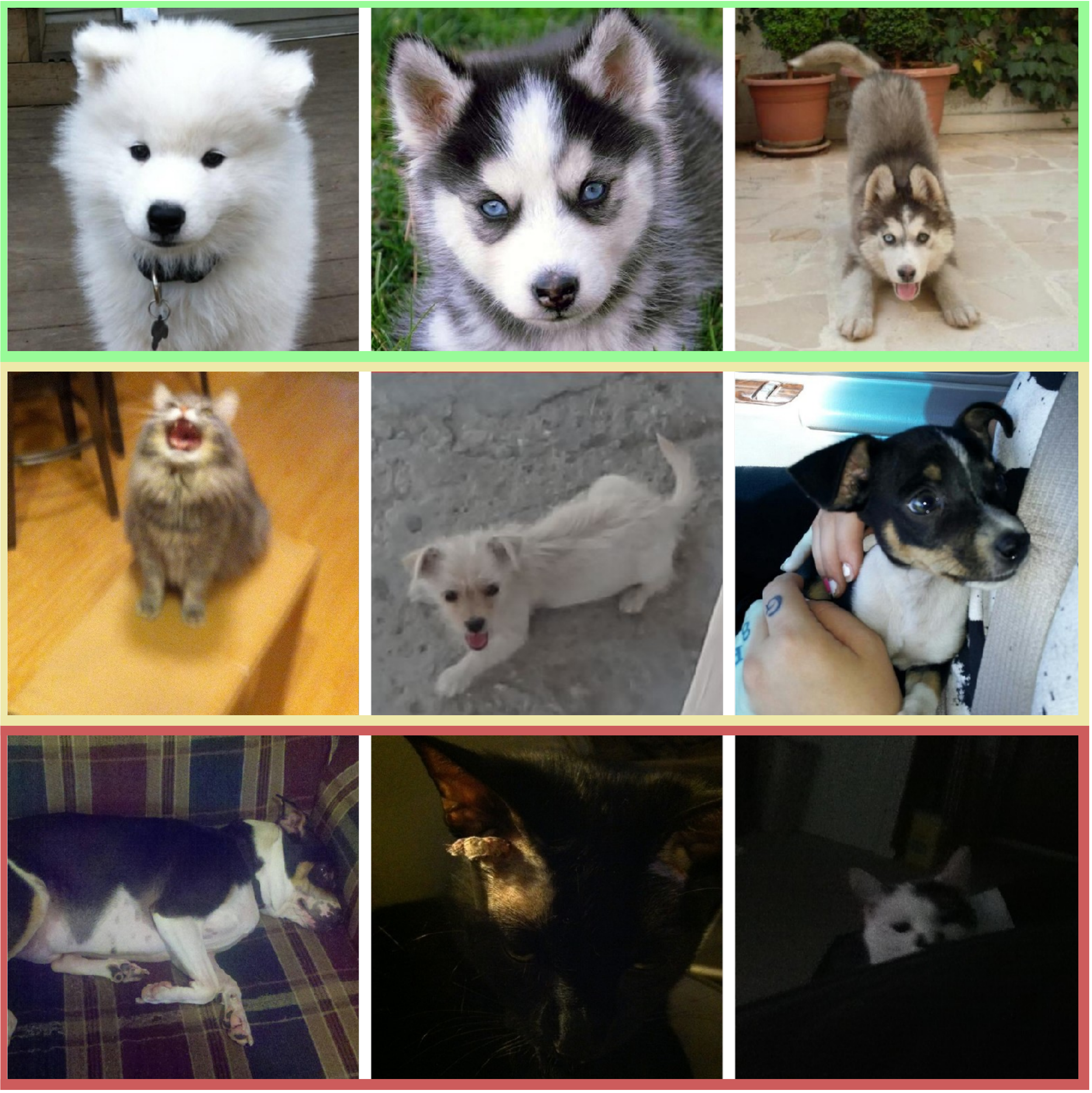}
  \caption{Image only}
  \label{fig:aww-img}
\end{subfigure}%
\begin{subfigure}{.33\textwidth}
  \centering
  \begin{tabular}{l}
    \input{aww-captions.tex}
  \end{tabular}
  \caption{Text only}
  \label{fig:aww-text}
\end{subfigure}
\begin{subfigure}{.33\textwidth}
  \centering
  \begin{tabular}{{C{2.5cm}C{2.5cm}}}
    \input{aww-multimodal.tex}
  \end{tabular}
  \caption{Multimodal}
  \label{fig:aww-mm}
\end{subfigure}
\caption{Examples from one train/test split of \communityname{aww}
  scored by the ResNet50 model, the unigram model, and the text
  \texttt{+} image model. The top, middle, and bottom rows are sampled
  from the 99th, 50th, and 1st percentiles respectively. Images
  courtesy \texttt{imgur.com}.
}
\label{fig:examples}
\end{figure*}

However, we understand that readers may still be curious to know
whether the ranking induced by our method has any correlation with the
scores that appear on Reddit, since other work
(e.g., Khosla et al. \cite{khosla2014makes}, who worked with a Flickr
dataset) computes similar correlations.
To satisfy the curious
reader, we did go ahead and compare the Spearman correlation between
raw popularity and our model's scores.  For the text + image models,
the observed values averaged over cross-validation splits range
between $\rho = .25$ for \communityname{pics} and $\rho = .37$ for
\communityname{MakeupAddiction}.
In general, the correlations we observe are somewhat lower than those
of Khosla et al.'s \cite{khosla2014makes} image-based model; whether
the differences are due to the models or to the different domains is
an open question.

\section{Analysis of aww}

We now qualitatively analyze the models' performance on
\communityname{aww}, though a similar analysis could be performed on
any community (e.g., Figure~\ref{fig:FoodPorn} shows image examples
from \communityname{FoodPorn}). Figure~\ref{fig:examples} shows
several test examples scored by the image-only, text-only, and
multimodal models from one of the \communityname{aww} cross-validation
splits.

Figure \ref{fig:aww-img}, which displays good, okay, and bad images as
scored by ResNet50, illustrates that lighting is
 important. The model tends to assign lower scores in cases
where an animal's face isn't visible. Having the animal taking up a
majority of the image also seems to be important, though this could be
an artifact of our resizing procedure. Also, we noticed that a
disproportionate number of highly scored images were of dogs; among
the cross-validation split we considered, in fact, the top ten images
were all dogs. The model, and potentially the community, might be
favoring particular types of animals.

To examine this possibility, we turn our focus to more interpretable
object detections. Specifically, we turn to the canonical 1K ImageNet
classes, which consist of a surprisingly high number of types of
animals, e.g., 120/1000 classes are different types of dogs. As such,
these classes are well-suited to analyzing \communityname{aww}.
We extracted the pre-softmax input for each ImageNet class according
to ResNet50 for each image \footnote{Weights after the softmax
  transformation also produced some significant results, but the
  pre-softmax weights are known to contain more fine-grained
  information \cite{bucilua2006model} } These features are the
un-normalized log probabilities for each of the 1K ImageNet
classes. For each of the 15 cross-validation splits, we computed the
average Pearson correlation between our model's score and the object
detection features.

After applying Bonferroni-correction to our confidence intervals to
account for the fact that there are 1K possible correlations, we
observed many significant results. Among the 250 most common
detections, the object-like features most correlated with success were
``golden retriever,'' ``dingo,'' and ``labrador retriever'' ($R =$
.23, .21, .19, respectively, $p \ll .01$ that there is a true
correlation). There were also dog breed features associated with
failure, including ``miniature schnauzer,'' ``maltese dog,'' and
``affenpinscher'' ($R=$ -.23, -.21, -.21, $p \ll
.01$). Interestingly, non-bulldog terriers fared poorly; all 15 were
negatively correlated with model score, though only 12/15 were
significantly so. In contrast, 5/5 retriever classes were
significantly correlated with higher scores. For cats, ``cheetah'' and
``lion'' features positively correlated ($R = .18, .09$) while
``tabby,'' ``egyptian cat,'' and ``persian cat'' features were all
negatively correlated ($R=$ -.1, -.11, -.17).

The story with text on \communityname{aww} is a simpler;
Figure~\ref{fig:aww-text} shows that longer captions generally do
better, and it also helps to have a story. Unigrams like saved
($\beta$ = .50), wife ($\beta$ = .43), roommate ($\beta$ = .42), and
cancer ($\beta=.41$), and are among the most predictive of
success. Interestingly, sleeping animals seem to be predictive of
failure, with unigrams like sleepy ($\beta$ =-.58), sleeping
($\beta$=-.47), laying ($\beta$=-.47), and nap ($\beta$=-.43) being
among the most predictive of failure.

When image and text features are combined, performance improves over
each by themselves, which suggests that the patterns discussed contain
information orthogonal for predictive purposes. Because we simply
concatenate image and text features rather than modeling interactions
directly, the multimodal patterns likely mirror the unimodal patterns
discussed here.

\section{Additional Related Work}

Content has been used to predict popularity in the past. Language
\cite{petrovic2011rt,hong2011predicting,Guerini:ProceedingsOfIcwsm:2011,bandari2012pulse,danescu2012you,artzi2012predicting,Sun+Zhang+Mei:13,tan2014effect,Tsur+Rappoport:12},
images \cite{khosla2014makes,Deza_2015_CVPR,wu2016time}, video
\cite{shamma2011viral,figueiredo2013prediction,pinto2013using}, or a
combination of multiple modalities
\cite{yamaguchi2014chic,mcparlane2014nobody,gelli2015image,hu2016multimodal,chen2016multi}
have been used for this task. Some previous work has controlled for,
rather than modeled, multimodal content
\cite{borghol2012untold,lakkaraju2013s}. Our work builds upon previous
studies that attempt to predict or analyze \emph{crowd-level}
preferences
\cite{khosla2014makes,figueiredo2014does,bakhshi2014faces,bakhshi2015red,stoddard2015popularity,schifanella2015image,Deza_2015_CVPR,mazloommultimodal,almgren2016predicting,zakrewsky2016item},
as opposed to \emph{user-level} preferences
\cite{zhong2015predicting}. After submission, we discovered a blog
post by Glenski and
Stoddard\footnote{\href{https://goo.gl/9M6Ioh}{\texttt{https://goo.gl/9M6Ioh}}}
describing human experiments similar to ours. While the setting we
examine is different (e.g., we apply more stringent time controls), it
was interesting to see that their human trial results were similar to
ours.

\paragraph{Noisy Rich-get-richer Processes.}
Timing \cite{borghol2012untold,lakkaraju2013s}, and even early random
positive or negative treatments \cite{weninger2015random} can affect
the popularity of social media content. Salganik et
al. \shortcite{salganik2006experimental} show that while content does
matter to an extent, presenting different orderings of songs to users
results in wildly different most and least popular music. These effects
likely underpin the widespread underprovision on Reddit
\cite{gilbert2013widespread}, which causes ``Reddit [to overlook] 52\%
of the most popular links the first time they were submitted.''
Undoubtedly, content can never perfectly predict community response.

\paragraph{Social Features for Eventual Popularity.}
Social connections \cite{lerman2010using} and author identity
\cite{suh2010want} also effect the popularity of content. Solomon and
Herman \shortcite{solomon1977status} demonstrate that individuals with
higher status are more likely to be recipients of prosocial
behavior. In our case, this could mean higher status individuals in a
community receive upvotes as a result of their celebrity
status. Khosla et al. \cite{khosla2014makes} consider a simple set of
social features of their Flickr dataset, and find that social features
are significantly more predictive of popularity than image features
when not controlling for user identity.

\section{Conclusion and Future Work}

In this work, we motivated the task of relative popularity prediction
as a means of controlling for time. We also demonstrated that
incorporating multimodal features generally resulted in improved
performance. Future work in modeling could consider more sophisticated
models of textual and visual interaction. Also, it would be
interesting to investigate visual trends within communities over
time. Designing a model to identify ``timely'' or trend-setting image
features is a promising avenue for future work.

Popularity prediction, too, is only one social factor of interest to
moderators of multimodal communities. The text of comments, for
example, offers a more fine-grained measure of community response than
upvotes. Text features like sentiment could also be predicted from
content in a similar time-controlled setting.

While we've provided evidence that there exist online communities
wherein visual and textual content predict popularity more
successfully than social features, it is important to point out the
results presented here might not generalize to other communities,
e.g., ones off of Reddit. We suspect that social connections are less
salient on Reddit, which seems more centered on the content.
Instagram, for example, is a social network based on image content
wherein identity likely matters more. However, even on Reddit itself,
we observed a case in \communityname{RedditLaqueristas} where our
intuitions proved to be incorrect: celebrity-status/social features
were more predictive than content in that subreddit.

Another caveat: while sampling pairs of posts made in quick succession
provided good timing/ordering controls for us, in other settings there
might not be enough posts to warrant such a sampling technique.

In the end, predicting what becomes popular in any given community
requires accounting for timing, content, identity, social structure,
and self-reinforcing rich-get-richer processes. While the relative
predictive power of each varies on a case-by-case basis, we hope the
results presented here encourage practitioners to investigate
content-driven models in the face of complex confounding factors.

\paragraph{Acknowledgments.} This material is based upon work
supported by the National Science Foundation under grant
no. IIS-1526155 and a Yahoo Faculty Research and Engagement Program
grant. We would like to thank Nvidia for the Titan X GPUs used in this
study, and Vlad Niculae for providing the website template used in the
human study. We also thank Yoav Artzi, Claire Cardie, Yiqing Hua, Vlad
Niculae, Tianze Shi, and the anonymous reviews for their helpful,
detailed feedback. We thank all members the Cornell NLP seminar for
their helpful comments. We thank B. Adler, J. Bass, A. Callahan,
G. Correa, E.  Danko, M. Feldman, D. Foster, M. Grusky, S. Helman, D.
Hessel, T. Hessel, A. Hirsch, L. James, V. Kern, E.  Krichilsky,
M. Magnusson, K. Mast, G. Pleiss, H. Rishi, L. Rosen, R. Rotabi,
A. Schofield, T. Shi, A.  Singh, A. Smith, A. Suhr, W. Thomason,
L. Thompson, M.  Zeich, and J. Zhang
for participating in the human study.

\bibliographystyle{abbrv}
\bibliography{refs-llee-macros,refs-llee}  %

\end{document}